\documentclass[pdflatex,sn-mathphys-num]{sn-jnl}% Math and Physical Sciences Numbered Reference Style
\usepackage{graphicx}%
\usepackage{multirow}%
\usepackage{amsmath,amssymb,amsfonts}%
\usepackage{amsthm}%
\usepackage{mathrsfs}%
\usepackage[title]{appendix}%
\usepackage{xcolor}%
\usepackage{textcomp}%
\usepackage{manyfoot}%
\usepackage{booktabs}%
\usepackage{algorithm}%
\usepackage{algorithmicx}%
\usepackage{algpseudocode}%
\usepackage{listings}%
\usepackage{graphicx}% Include figure files
\usepackage{dcolumn}% Align table columns on decimal point
\usepackage{xcolor}
\usepackage{bm}% bold math
\usepackage{physics}
\usepackage{breqn}
\usepackage{ulem}
\usepackage{hyperref}
\hypersetup{
	colorlinks = true,
	urlcolor   = blue,
	linkcolor  = blue,
	citecolor  = red
}
\usepackage{amsmath}
\usepackage{algorithm}
\usepackage{algpseudocode}
\usepackage{float}
\usepackage{siunitx}
\theoremstyle{thmstyleone}%
\theoremstyle{thmstyletwo}%

\theoremstyle{thmstylethree}%

\begin{document}

\title[Article Title]{An experimental investigation of quantum frequency correlations resilience against white and colored noise}

%%=============================================================%%
%% GivenName	-> \fnm{Joergen W.}
%% Particle	-> \spfx{van der} -> surname prefix
%% FamilyName	-> \sur{Ploeg}
%% Suffix	-> \sfx{IV}
%% \author*[1,2]{\fnm{Joergen W.} \spfx{van der} \sur{Ploeg} 
%%  \sfx{IV}}\email{iauthor@gmail.com}
%%=============================================================%%

\author*[1]{\fnm{Linda} \sur{Sansoni}}\email{linda.sansoni@enea.it}

\author[1]{\fnm{Eleonora} \sur{Stefanutti}}\email{eleonora.stefanutti@enea.it}

\author[1]{\fnm{Andrea} \sur{Chiuri}}\email{andrea.chiuri@enea.it}

\affil[1]{\orgdiv{Nuclear Department}, \orgname{ENEA}, \orgaddress{\street{Via E. Fermi 45}, \city{Frascati}, \postcode{000100}, \country{Italy}}}

%%==================================%%
%% Sample for unstructured abstract %%
%%==================================%%

\abstract{Understanding the impact of disturbances in quantum channels is of paramount importance for the implementation of many quantum technologies, as noise can be detrimental to quantum correlations. Among the various types of disturbances, we explore the effects of white and colored noise and experimentally test the resilience of a quantum ghost spectrometer against these two types of noise, showing that it is always robust against white noise, whereas colored noise introduces a huge impact on the process.}


%%\pacs[JEL Classification]{D8, H51}

%%\pacs[MSC Classification]{35A01, 65L10, 65L12, 65L20, 65L70}

\maketitle

\section{Introduction}\label{sec1}

Studying the propagation of quantum light in turbulent and noisy media is crucial for advancing our understanding of fundamental physics and improving technological applications. Disturbances can significantly affect the coherence, direction, and intensity of light, posing challenges in quantum science and its applications  \cite{cler10rmp}. Understanding how quantum correlations interact with such media can lead to breakthroughs in quantum communication, computation and sensing, where precision and reliability are paramount \cite{Karpinski2020_AQT, Lib2022_NatPhys, Mukamel2020}. Indeed, this research provides valuable insights into the fundamental principles of quantum mechanics and plays a central role in the advancement of new quantum technologies that can harness quantum advantage, even in highly unpredictable conditions.

The main fields potentially affected by this detrimental aspect include imaging \cite{padgett17ptrsa}, microscopy \cite{Bowe23conph} and spectroscopy \cite{Jana18prap}, where the disturbance-induced fluctuations, due to absorption, scattering, and variations in the refractive index could result in a loss of the quality of the expected outcomes. In recent years, there has been growing interest in understanding and mitigating the impact of disturbances on quantum technologies operating in real-world free-space channels, often affected by turbulence and noise. Nevertheless, most of the efforts have been focused on experimental and theoretical studies regarding quantum imaging and correlation-based imaging \cite{Morris15,John22opcon}. 
These techniques have proven effective in overcoming challenges associated with traditionally difficult free-space channels, e.g. ghost imaging (GI) in turbulent channels \cite{Hardy11,Dixon11,Shi12,Freitas25,Li2010}, quantum illumination in noisy environments \cite{Gregory20} or chaotic-light correlation imaging against turbulence \cite{Scala2024}. 
Several methods related to the \textit{denoising} of quantum images have also been developed and tested via imaging distillation \cite{Fuenzalida23},
image compression and reconstruction \cite{Morris15} or heralding of correlated photon pairs \cite{John22opcon,10.1063/5.0016106}.

However, the same attention has not been devoted to ghost spectroscopy (GS) and, more in general, to the frequency domain, as only a few examples are reported in the literature, all of which are based on environmental noise artificially introduced into the experimental setup \cite{Sanna24}.

In this study, we aim at assessing the resilience of quantum correlations under different levels of disturbance. In particular, we consider white and colored noise. White noise arises from quantum fluctuations in the field that are evenly distributed across frequencies, such as in a vacuum state or classical thermal noise in detectors. In contrast, colored noise exhibits frequency-dependent variations, typically induced by environmental coupling or quantum interactions with a thermal bath.
We depict a novel approach to experimentally engineer spectral disturbance characterized by specific features, corresponding to either white or colored noise.
Specifically, our investigation is built upon the framework of quantum ghost spectroscopy (QGS) \cite{chiu22acsph}, initially introduced as the counter-part in the frequency domain of the quantum ghost imaging \cite{Pittman95}, as a relevant testbed to prove the effects of different kinds of noise on quantum correlations encoded in the frequency domain.\\
The GS has been theoretically and experimentally studied with classical (thermal) light \cite{Jana18prap} and quantum sources of photon pairs based on spontaneous parametric down conversion (SPDC) \cite{chiu22acsph,Chiu25epj+} and spontaneous four-wave mixing \cite{Sanna24}. 
Significant research efforts have also focused on comparing these two approaches, highlighting the differences and similarities between quantum and classical light sources \cite{bennink04prl,sullivan10pra,Chiu22pra, Gatti04}. The possibilities offered by the quantum regime have been deeply explored developing highly non-degenerate SPDC sources to link the mid-infrared (MIR) and visible / near-infrared (NIR) range \cite{Yabushita04,Chan09,Aspden15,Tashima24,Neves24}. The frequency correlations characterizing these photon pairs have allowed to prove the feasibility of nonlinear microscopy \cite{Aspden15,Samantaray2017,Kalashnikov16,Gili22}, quantum spectroscopy  \cite{Lindner20,Mukai21,Kaufmann22,Tashima24,Kurita25} and quantum holography \cite{Grafe22}.\\
\indent Our work represents the first experimental realization of fully controlled white and colored noise encoded in the frequency domain to assess the performances of QGS in a highly non-degenerate configuration. We demonstrate that the QGS and the quantum correlations are always resilient against white noise, while colored noise introduces a significant impact on the system. These findings provide experimental evidence that frequency correlations can be made robust against noisy backgrounds in free-space channels by suitably engineering and adapting the photon-pair source. Additionally, the reduction of colored noise, arising from accidental coincidence counts between uncorrelated photons, enables operations at higher brightness, which in turn allows for video-rate acquisition and more realistic integration times \cite{McCarthy25, Fisher-Levine_2016,NOMEROTSKI201926,Vidyapin2023}.\\
\indent The article is organized as follows. In Section 2, we describe the approach adopted to simulate white and colored noise in QGS, we then present our experimental setup in Section 3 and the results in Section 4. In Section 5, we discuss our findings and comment on the resilience of quantum frequency correlations against noise. Finally, in the conclusive section, we provide perspectives on potential generalizations and future extensions of the present work.

\section{Methods}
\begin{figure}[h!]
   \centering
   \includegraphics[width=0.8\columnwidth]{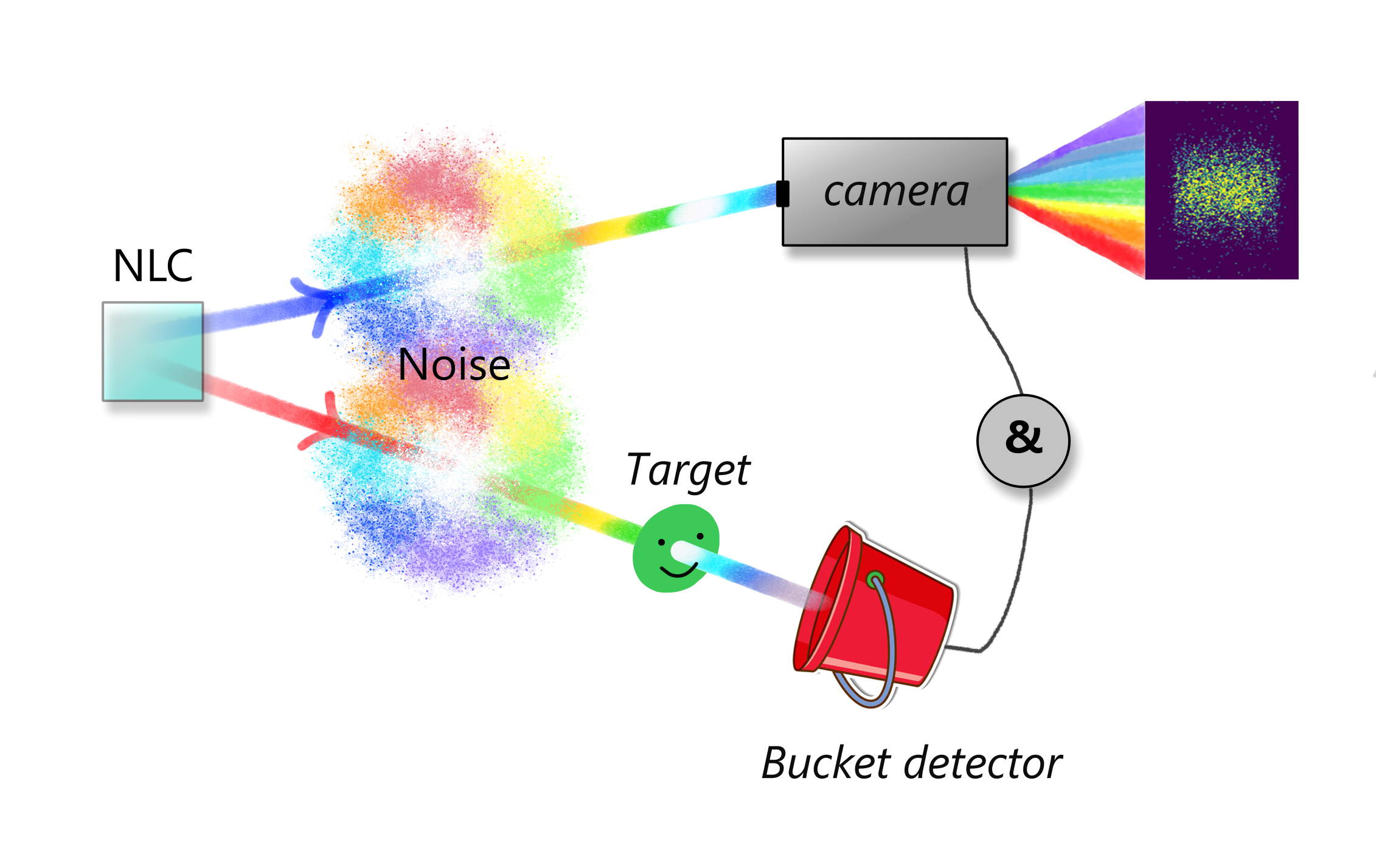}
   \caption{Conceptual scheme of the experiment: photon pairs from a non-linear crystal (NLC) disturbed by either white and colored noise are used to perform a QGS experiment \cite{chiu22acsph}.}
   \label{fig:setup}
\end{figure}

We use the QGS as a platform to test the effects of noise on quantum correlations. QGS is based on the frequency correlations of two-photon pairs generated through SPDC in a non-linear crystal \cite{Boyd20bk}. A conceptual scheme of our analysis is shown in Fig. \ref{fig:setup}: here one photon from a down-converted pair in a non-linear crystal is sent on a target before being revealed by a bucket detector, the other one is analyzed in the spectral degree of freedom, and coincidences between these two photons are collected. Both channels are affected by either white or colored noise.
Since SPDC through a non-linear crystal is the underlying process of our experiment, we consider the following interaction Hamiltonian \cite{wall08bk}:

\begin{equation}
    \mathcal{H}_{int}=i\hbar\chi E_P[\hat{a}^{\dagger}_s\hat{a}^{\dagger}_i+\hat{a}_s\hat{a}_i],
\end{equation}
where $\hat{a}^{\dagger}_j$ ($\hat{a}_j$) for $j=s,i$ is the creation (annihilation) operator for photons in the signal and idler modes, respectively, $\chi$ is related to the nonlinearity of the crystal and $E_p$ is the amplitude of the pump electric field.
The corresponding evolution operator is then:
\begin{equation}
    U=e^{-\frac{i\mathcal{H}t}{\hbar}}=\exp{\left[\Gamma(\hat{a}^{\dagger}_s\hat{a}^{\dagger}_i+\hat{a}_s\hat{a}_i)\right]}
    \label{eq:U_spdc}
\end{equation}

\noindent
where $\Gamma$ is the non-linear gain.
By using a Taylor expansion of the second order and considering a disentangling theorem \cite{ross02bk}, the operator in Eq. (\ref{eq:U_spdc}) applied to the vacuum state can be rewritten as:
\begin{eqnarray}
\ket{\Psi}=U\ket{0_s,0_i}&&=\frac{1}{C}\left[1+\Gamma\,\hat{a}^{\dagger}_s\hat{a}^{\dagger}_i+\frac{\Gamma^2}{2!}(\hat{a}^{\dagger}_s\hat{a}^{\dagger}_i)^2\right]\ket{0_s,0_i} \\
&&=\frac{1}{C}\left[\ket{0_s,0_i}+\Gamma\,\ket{1_s,1_i}+\frac{\Gamma^2}{2!}\ket{2_s,2_i}\right]\nonumber
\end{eqnarray}
where $C$ is a suitable normalization constant. The last expression clearly shows that the SPDC process generates either one or two photon pairs in the signal and idler modes. These modes can be defined by various degrees of freedom: spatial mode, polarization, frequency, etc. In our case, we deal with a source of frequency-entangled pairs, meaning that our signal and idler modes correspond to two distinct frequencies. Nevertheless, we have now to take into account that signal and idler photons have a spectral distribution, thus the non-vacuum part of the state generated by the SPDC, up to a normalization constant, is rewritten as
\begin{eqnarray}
        \ket{\Psi}=&&\Gamma\int d\omega_sd\omega_if(\omega_s,\omega_i)\hat{a}^{\dagger}_{\omega_s}\hat{a}^{\dagger}_{\omega_i}\ket{0_s,0_i}\label{eq:spdc_freq1}\\
        +&&\frac{\Gamma^2}{2!}\int d\omega_sd\omega_i\omega_s^{\prime}d\omega_i^{\prime}f(\omega_s,\omega_i)f(\omega_s^{\prime},\omega_i^{\prime})\hat{a}^{\dagger}_{\omega_s}\hat{a}^{\dagger}_{\omega_i}\hat{a}^{\dagger}_{\omega_s^{\prime}}\hat{a}^{\dagger}_{\omega_i^{\prime}}\ket{0_s,0_i}\nonumber
    \end{eqnarray}
Here $f(\omega_s,\omega_i)$ is the joint spectral amplitude (JSA) of the generated photon pairs at different frequencies $\omega$, which contains information about the frequency distribution of the pair. While the first term of eq. (\ref{eq:spdc_freq1}) describes the generation of a frequency correlated pair, the second term contains the product of two independent JSAs, corresponding to the generation of two independent pairs of photons.
When measuring only coincidences between two photons, as in the case of the experiment depicted in Fig. \ref{fig:setup}, the second term in eq. (\ref{eq:spdc_freq1}) gives rise to the detection of uncorrelated two-photon events, corresponding to those events where the two detected photons belong to different pairs \cite{Sanna24, Signorini2020}.
Since the JSA function of photon pairs is the same, this second contribution has the same spectral distribution as the correlated events described by the first term in (\ref{eq:spdc_freq1}), thus it can be considered as colored noise.
Indeed this effect introduces a disturbance in the frequency domain: noisy photons from two different emitted pairs affect the process by introducing uncorrelated events, whose frequencies overlap with the correlated ones.
The probability of generating more than one pair is proportional to the non-linear gain $\Gamma^2$ and increases with the power of the pump laser (given by $|E_p|^2$).
This contribution is detrimental to any process based on correlations. In particular, when exploiting frequency entanglement of SPDC pairs, the introduction of uncorrelated pairs lowers the degree of correlation between the generated photons, leading to worst performances of any protocol relying on such correlations. Moreover, when the pump laser is pulsed, this effect is further influenced by events due to coincidences between photons generated at two different times, corresponding to two consecutive pulses of the pump laser.

Noise can be minimized by adjusting experimental parameters. In particular, colored noise in SPDC due to double-pair generation can be minimized by operating in a low-rate generation regime; however this results in low count rates and, in turn, long measurement times.
Assuming time-consuming measurements can still be acceptable, working at very low pump powers can completely avoid colored noise. Nevertheless, the process remains not completely noise-free. Indeed, the decreasing of the generation rate introduces other background noises, in particular spurious counts from the detection devices.
{This last effect is related to the main parameters commonly used to characterize the performance of a detection system:}\\
{- Dark count rate (for single-photon avalanche diode SPAD) or dark current (in the CCD camera) and electron background illumination (of the intensifier): represent signal detection when the sensor is in the dark, primarily due to thermal generation of free carriers, and it determines the minimum incident photon rate that can be detected. This parameter is relevant when the detected signals are close to these thresholds.\\}
%{- Detection efficiency: it is the ratio of the number of pulses detected to the number of total incident photons.\\}
{- Afterpulsing probability: during detection, some carriers could trigger a detection cycle not initiated by photon absorption. These non-photon-triggered pulses are referred to as afterpulsing. }\\
{- Dead time: it refers to the period after each detection event during which the detector is unable to record another event. Setting short dead times allows for higher count rates, however it may also increase afterpulsing events, as well as dark counts.}

The spurious contributions due to the aforementioned phenomena can be modeled as white noise.
White noise is generally harder to reduce through correlation techniques \cite{paz16pra,wang24arx,zou24npjqi,mass22epj+}, but it can often be mitigated by averaging over large datasets or using filters designed to reduce specific types of noise.

A figure of merit often used to evaluate the noise contribution of a setup is the signal-to-noise ratio (SNR), which, in our case, is given by the coincidence-to-accidental ratio (CAR) \cite{Guo17ape,Kuma21apl,Hojo23scirep}.
It can be calculated as:
\begin{equation}
   CAR = \frac{N_{cc}}{N_{acc}}, 
\end{equation}
where $N_{cc}$ is the raw two-photon coincidence counts, while $N_{acc}$ denotes the accidental coincidence counts (uncorrelated photon pairs).
Usually, experiments based on correlations are performed with a CAR value as high as possible. 
At low pump power, the CAR results to be limited by dark counts and environmental noise (white noise). Conversely, at higher pump powers, the CAR is limited by after-pulsing and multi-pair generation (colored noise), which increases the probability of accidental coincidence counts between signal and idler photons belonging to different pairs. 

Depending on the process under consideration and the maximum acceptable noise, the parameters of an experiment are typically defined according to CAR value.
However, by investigating the effects of white and colored noise in the frequency degree of freedom of a QGS experiment, we demonstrate that the CAR alone is not a sufficient figure of merit to assess whether the protocol is noise-resilient. A crucial factor is the knowledge about the nature of the disturbance. Indeed, we observe that for low values of the CAR the QGS is resilient against white noise, while it is vulnerable against colored noise.
{It is important to note that the amount of white and colored noise depends on the setup equipment and parameters. It is important to note that the amount of white and colored noise depends on the setup equipment and its parameters. Therefore, a comprehensive characterization can be carried out before the experiment in order to find the optimal settings. In our case, prior knowledge allowed us to fix some parameters at their optimal value - such as detector dead time, source brightness, coincidence time-window- while  varying others that we found most relevant to our investigation.

\section{Experiment}
Our experimental strategy relies on the nonlinear process of SPDC, wherein a small fraction of photons from a pump laser is converted into correlated photon pairs with lower energies. Our experiment was conducted in a non-degenerate configuration, where the signal and idler down-converted photons were in the near-infrared and visible spectral range, respectively. The QGS approach exploits the intrinsic frequency correlations between the generated signal and idler photons to enable spectral characterization of infrared photons by performing measurements on their visible counterparts \cite{chiu22acsph}.
\begin{figure}[h!]
   \centering
   \includegraphics[width=\columnwidth]{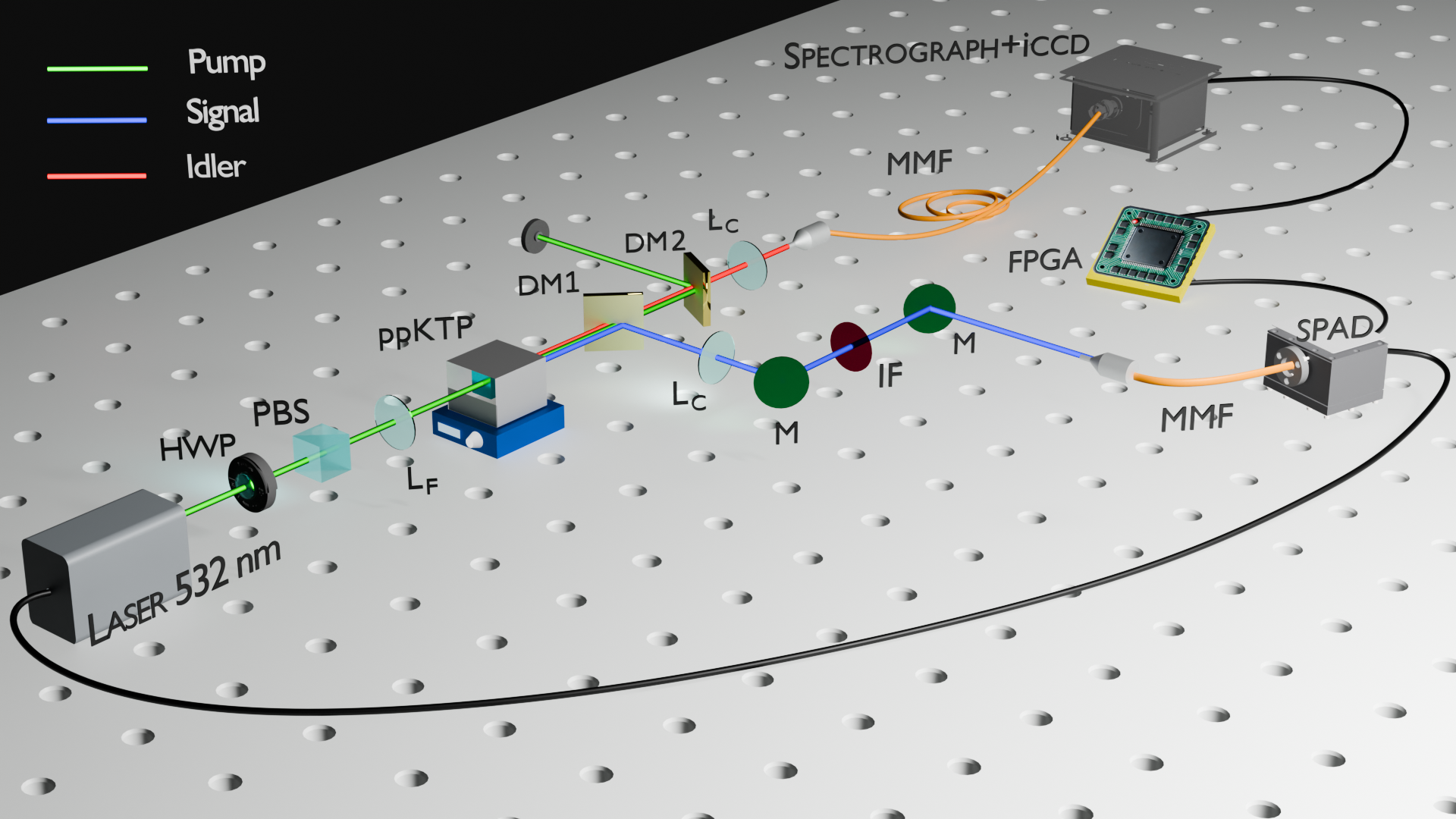}
   \caption{Experimental setup for QGS measurements. The optical elements in figure are HWP: half-wave plate, PBS: polarizing beam splitter, ppKTP: non-linear crystal, DM1: dichroic mirror for separating signal (reflected arm) and idler (transmitted arm) photons, DM2: dichroic mirror used to remove the pump beam from the idler path, IF: interference filter, L$_{\rm{F}}$: focusing lens (f = $250\,\rm{mm}$), L$_{\rm{C}}$: collimating lens, M: mirror, MMF: multi-mode fiber.}
   \label{fig:setup2}
\end{figure}

\noindent Fig. \ref{fig:setup2} illustrates a schematic of our QGS experimental setup. The excitation beam was generated by a pulsed laser operating at a repetition rate of $40\,\rm{MHz}$ with a pulse duration of $8\,\rm{ps}$; the pump spectrum was centered at $\lambda_P = 532\,\rm{nm}$, with a spectral bandwidth of $0.23\,\rm{nm}$. The laser beam was used to illuminate a $2$-mm-long periodically poled potassium titanyl phosphate (ppKTP) crystal (manufactured by SLF Svenska Laserfabriken AB), with a poling period of $9.725\,\rm{\mu m}$. The type-0 down-conversion process generated pairs of collinear photons with vertical polarization, matching that of the pump laser.
The phase matching condition was optimized with the crystal temperature, which could be tuned using a customized thermoelectric heating cell and a temperature controller with a precision of $0.1^\circ \rm{C}$. The operating temperature was kept at $T=30^\circ\rm{C}$, at which the signal photons are emitted at a central wavelength $\lambda_i \approx 1550\,\rm{nm}$, corresponding to an idler wavelength $\lambda_s \approx 810\,\rm{nm}$. 
The idler and signal photons were separated using a dichroic mirror (DM1). The signal photon on the reflected arm of DM1 was directed to an interference filter centered at $1550\,\rm{nm}$ with a bandwidth of $10\,\rm{nm}$ and then coupled to a single-photon avalanche diode (SPAD, MPD PDM-IR), acting as a bucket detector. The SPAD was triggered by the sync-out TTL (transistor-transistor logic) signal coming from the laser.
In order to perform frequency-resolved measurements, the idler photon on the transmitted arm of DM1 was sent to a spectrograph (Andor Kymera 328I-A-SIL) equipped with a diffraction grating (600 lines/mm) and an intensified CCD (Andor iSTAR iCCD DH334T-18U-73). Both idler and signal photons were collimated and subsequently collected through multi-mode fibers. The pump laser was removed by a second dichroic mirror (DM2) along the idler path.\\
To measure signal-idler photon coincidences, the acquisition from the camera was synchronized with the SPAD, taking into account optical path length differences between signal and idler photons, the bucket detector response time, and the other electronic delays. Thus the correct delay on the idler was introduced using a multi-mode optical fiber of appropriate length and finely adjusted with a field programmable gate array (FPGA) board connected to the SPAD detector.
The raw data acquired from our measurements represent the photon counts recorded by the iCCD for each pixel in photon-counting mode. Since the horizontal coordinate of each pixel corresponds to a wavelength, we obtained a spectral analysis of the idler photons detected in coincidence with the signal counterpart recorded by the bucket detector. As the spatial distribution of the collected idler photons was not of interest, an integration over the vertical coordinate (vertical binning) was performed to obtain the total photon counts along with their spectral distribution. All final spectra were smoothed using a Savitzky-Golay filter. Given the idler spectrum, the corresponding signal spectral distribution was inferred as a consequence of the phase-matching condition ($\displaystyle \lambda_s = (1/\lambda_P -1/\lambda_i)^{-1}$).

By properly adjusting the delay between iCCD and SPAD, we collected photon counts corresponding to signal-idler coincidences and accidental coincident counts due to uncorrelated photons. This allowed us to perform the ghost measurements of the interference filter and estimate the coincidence-to-accidental ratio of our source as a function of the pump power. The average power of our laser was tunable up to $200\,\rm{mW}$, and the CAR dependence on pump power was characterized over the entire range. To obtain lower CAR values we have inserted a lens before the crystal to focus our pump beam and evaluated the CAR as a function of the pump power normalized to the beam area, i.e. pump density power.
The integration time was optimized for each measurement to ensure statistically significant photon counts across varying the pump power.  
As a reference, we measured the spectral properties of the source with the $2$-mm ppKTP non-linear crystal removing the IF from the signal path, setting the pump power at an %$10\,\rm{mW}$ in order to have the 
optimal CAR value.

\section{Results}\label{sec2}

In Fig. \ref{fig:noisy_regimes} we present an analysis of the noise in relation to the CAR of our system. Panel (a) displays the measured CAR as a function of the pump density power.
\begin{figure*}[t]
    \centering
    \includegraphics[width=\textwidth]{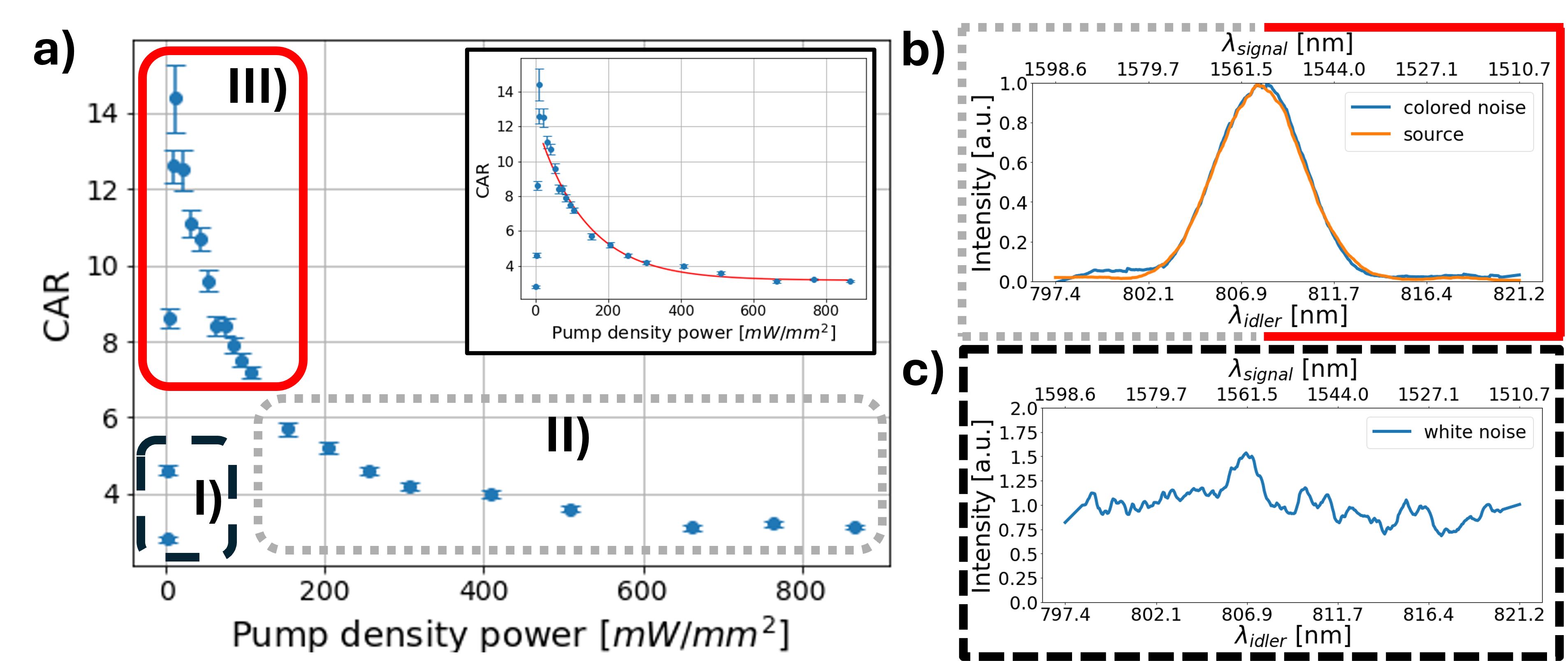}
    \caption{(a) CAR as a function of the pump density power. For values of CAR above $\rm{CAR}_{min}$ noise is negligible (box III), for values below $\rm{CAR}_{min}$ two regions are highlighted corresponding to two types of noise: white noise (box I) and colored noise (box II). The corresponding spectra for colored (b) and white (c) noise. Here the error bars have been calculated propagating the poissonian fluctuations on the photon counts. 
    }
    \label{fig:noisy_regimes}
\end{figure*}
The data show a steep rise, followed by a peak and a slow decrease. The decreasing trend (panel (a), top right inset) has been fitted  with a negative exponential function of the type $\rm{CAR}(P)=Ae^{-\frac{P}{P_o}}$, where $A$ is the maximum CAR and $P_o$ denotes the pump density power at which the CAR is lowered by a factor $e$. Using the fitting parameters, the CAR corresponding to $P=P_o$ has a value of $\rm{CAR}_{min}=(6.5 \pm 0.1)$, that can be interpreted as the minimum CAR threshold above which the impact of noise on the experiment becomes negligible (Fig. \ref{fig:noisy_regimes}(a) box III). Although the $\rm{CAR}_{min}$ value depends on the specific equipment adopted for the setup, its limit is indirectly taken into account in various experiments, where the parameters are adjusted to remain far enough from this threshold to ensure successful experiments \cite{Pitsch:21}. As a matter of fact, for CAR values below $\rm{CAR}_{min}$, noise becomes dominant. At low CAR two different regions, corresponding to two different types of noise, can be identified: one where the noise is due to white events, such as dark counts (panel (a) box I). Measured noise as a function of the signal and idler wavelengths corresponding to this region is shown in Fig. \ref{fig:noisy_regimes} (c), where a flat spectrum with random fluctuations is observed. In the second region (panel (a) box II) the noise is introduced by colored events, whose spectral distribution is shown in Fig. \ref{fig:noisy_regimes} (b). Here, both the source spectrum (orange curve) and the noise spectrum (blue curve) are shown. The two distributions are clearly the same, meaning that the noise presents spectral features that overlap with the process under investigation. As a result, the noise introduces disturbances that may be detrimental, as we will demonstrate later.

\begin{figure*}[t!!]
    \centering
    \includegraphics[width=\textwidth]{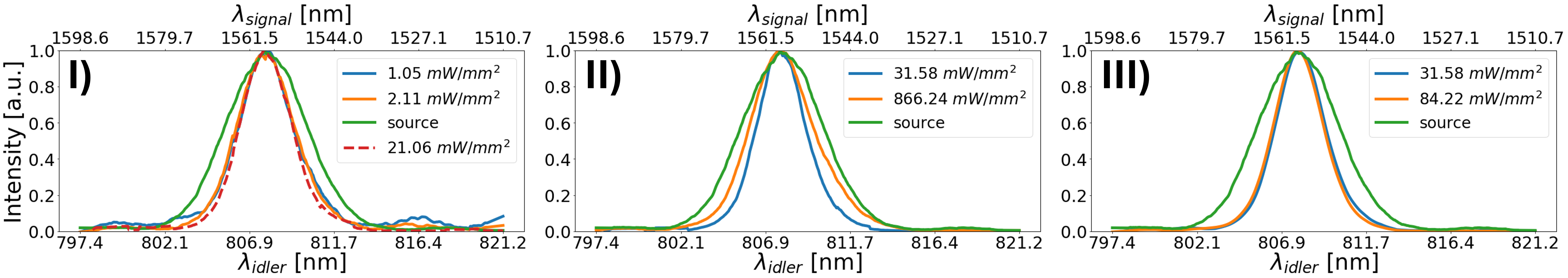}
    \caption{Comparison between the experimental spectra of the source and the IF ghost at different CAR values corresponding to Fig. \ref{fig:noisy_regimes} (a). I) Quantum correlations are robust in the presence of white noise and the QGS is not affected by it. II) In the presence of colored noise quantum correlations are fragile and broadening of the spectrum is observed. III) For optimal values of the CAR quantum correlations are preserved and the QGS measurements are fast and reliable.
    }
    \label{fig:broadening_panel}
\end{figure*}

\noindent Obviously, when setting up an experiment, the optimal working-point corresponds to a CAR as high as possible, i.e. on the peak of Fig. \ref{fig:noisy_regimes} (a). However, selecting a pump power density within the range corresponding to box III in panel (a) can still be advantageous, depending on other factors such as signal rates and measurement times.\\
After characterizing the CAR and noise spectrum, we performed a QGS experiment with the aim of reconstructing the spectral image of an interference filter (full width at half maximum, FWHM=10 nm) at different CAR values, corresponding to the three different regions of Fig. \ref{fig:noisy_regimes}: one at low CAR dominated by white noise (panel I), one at low CAR dominated by colored noise (panel II) and the last at the optimal CAR (panel III). The results are shown in Fig: \ref{fig:broadening_panel} panels I, II and III, respectively.

\begin{figure}[h]
    \centering
    \includegraphics[width=0.8\columnwidth]{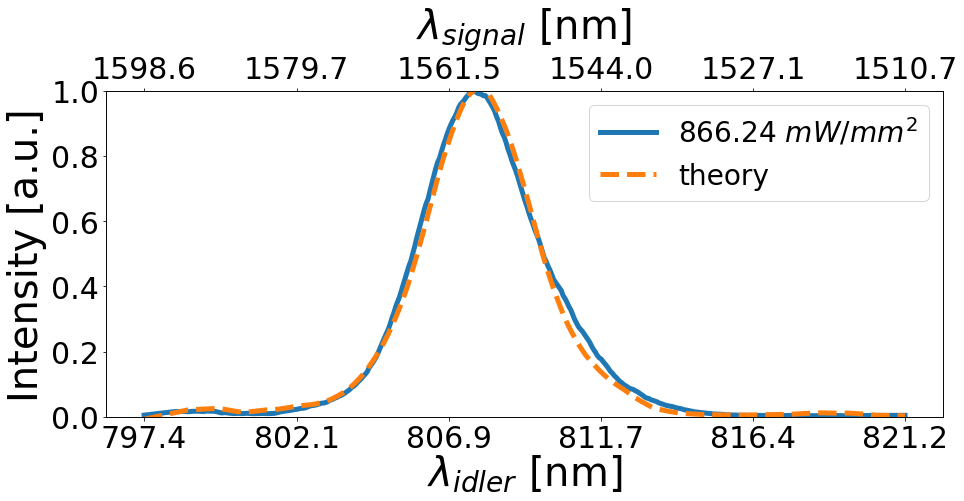}
    \caption{Comparison between experimental quantum ghost spectrum in presence of $30\%$ colored noise (blue curve) and the theoretical simulation (orange dashed line).}
    \label{fig:broadeningTeo}
\end{figure}

\section{Discussion}

Our results characterize the effect of white and colored noise on quantum correlations, in particular we observe that colored noise lowers the resolution of the QGS. This is very clear from Fig. \ref{fig:broadening_panel}: in panel I we report the ghost spectrum of the target IF taken at low pump density powers in comparison to the spectrum of the source (green line). For these values of pump density power the CAR is low and the noise relevant, but it is mainly due to white events. At the cost of long measurement times, it is nevertheless possible to reconstruct the ghost spectrum without alterations, meaning that the QGS (and thus quantum correlations) is robust against white noise. On the other hand, by increasing the pump density power, but still having a low CAR (Fig. \ref{fig:broadening_panel} II) colored noise plays a crucial role: as the pump density power increases, we observe a broadening of the ghost spectrum which tends to the width and shape of the source spectral distribution. In this condition the QGS protocol is no longer reliable, as its resolution is heavily reduced: here quantum correlations are strongly affected by colored noise and the quantum protocol fails. The optimal parameters - for CAR above $\rm{CAR}_{min}$ - lead to an optimal QGS protocol, whose results are shown in fig. \ref{fig:broadening_panel} III: the ghost spectrum of the target IF is retrieved without broadening and with high signal rates.

To ascertain that the broadening is due to the presence of colored noise, we have compared the experimental ghost of Fig. \ref{fig:broadening_panel} II (blue spectrum), corresponding to a noise of $30\%$, with a theoretical model (orange dashed line). The results are shown in Fig. \ref{fig:broadeningTeo}: the two shapes overlap, showing that the loss of resolution can be ascribed to the presence of colored noise.

For a better understanding of the dramatic effect of colored noise on the resolution of a QGS, we performed a simulation of the resolving power of two peaks (FWHM = 10 nm @1550 nm corresponding to FWHM = 2.8 nm @810 nm) by varying their spectral distance and the amount of colored noise.
We modeled the two peaks as Gaussian shapes and the resolving power as
\begin{equation}
    R_P=\frac{x_0^1-x_0^2}{\sigma_1+\sigma_2}
\end{equation}
where $x_0^{1(2)}$ is the central wavelength of the first (second) peak and $\sigma_{1(2)}$ are their respective standard deviations, or half widths. The two peaks can be resolved as long as $R_P>1$.
In Fig. \ref{fig:simul2peak} we show the behavior of $R_P$ as a function of the distance between the central wavelengths - varying between 0.5 and 3 nm for the idler and between 1.8 and 11.3 nm for the signal - and the amount of colored noise (with the same spectral distribution of the source, which is much larger than the combined spectral distribution of the two peaks). We can clearly observe that the resolving power dramatically decreases as soon as noise increases and, as expected, this effect becomes more pronounced when the peak separation is smaller. Nonetheless, even for two peaks that are 3 nm apart, an amount of noise around $25-30\%$ is sufficient to prevent their resolution.
\begin{figure}
    \centering
    \includegraphics[width=\columnwidth]{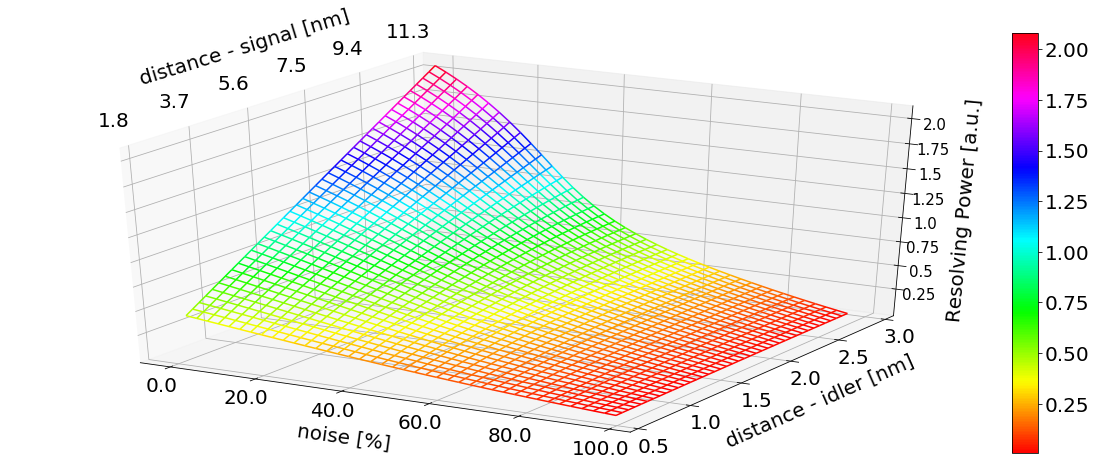}
    \caption{Resolving power $R_P$ as a function of the distance between the central wavelengths and the amount of colored noise. Here We have considered a spectrum composed by two peaks equal to the one obtained with Power Density = $31.58\,\rm{mW/mm^2}$. In this system we have introduced different percentages of colored noise repeating the simulation for different peak separations. Each resulting noisy spectrum was fitted using two Gaussian curves of the form $d+a\cdot \exp(-(x-x_0)^2/(2 \cdot \sigma^2))$. 
    $R_P$ has been calculated by using the parameters obtained from these Gaussian fits, where, $R_P > 1$ ($< 1$) means the two peaks can (cannot) be separated.}
    \label{fig:simul2peak}
\end{figure}
\section{Conclusions}
In this manuscript, we have provided an experimental investigation regarding the implementation of white and different spectral noises and their effect on quantum correlations. We have shown that the CAR value is not sufficient to determine the effect of noise on quantum correlations, as the nature of the noise itself plays a crucial role: quantum correlations are robust against white noise, while decoherence becomes relevant in the presence of colored noise. Our study is grounded in a platform that is attracting a broad interest for its possible applications and promising perspectives towards the 3D and remote sensing, i.e. the QGS. Technologies based on the \textit{ghost} approach rely on the use of two types of correlations: semiclassical, i.e. where a source of thermal or pseudo-thermal light is employed, or quantum. Since the latter has shown better performances \cite{bennink04prl,Chiu22pra}, we have explored this field. Recent experimental advances have pushed the capabilities of correlation-based technologies far beyond the expected boundaries thanks to their demonstrated advantages related to the visibility, the contrast-to-noise ratio and the resolution. Their performances can be extended and improved employing also higher-order correlations between more particles, as demonstrated with quantum and classical correlations \cite{PhysRevLett.122.233601,Chan:09}. These promising perspectives towards a realistic employment underscore the necessity of the systematic study presented in this work, which aims at a better understanding on how to proceed on the road to reliable quantum technologies.

\section*{Declarations}

\bmhead{Availability of data and materials} The datasets used and analyzed during the current study are available from the corresponding author on reasonable request.
\bmhead{Competing interests} The authors declare that they have no competing interests.
\bmhead{Funding} This project is funded within the QuantERA II Programme (Qucaboose Project), that has received funding from the EU H2020 research and innovation programme under GA No 101017733, and with funding organization NQSTI (Italy)
\bmhead{Authors' contribution} Conceptualization, Methodology, Experiment, Formal Analysis and Visualization, Writing, Review, and Editing: All Authors; Funding acquisition: A.C.
\bmhead{Acknowledgements} The Authors thank Max Widarsson and Patrick Mutter from SLF Svenska Laserfabriken for providing the source crystal, Federico Angelini for the fruitful discussions, and Marcello Nuvoli for the fabrication of mechanical holders.

%%===========================================================================================%%
%% If you are submitting to one of the Nature Portfolio journals, using the eJP submission   %%
%% system, please include the references within the manuscript file itself. You may do this  %%
%% by copying the reference list from your .bbl file, paste it into the main manuscript .tex %%
%% file, and delete the associated \verb+\bibliography+ commands.                            %%
%%===========================================================================================%%

\bibliography{biblio.bib}% common bib file

%% if required, the content of .bbl file can be included here once bbl is generated
%%\input sn-article.bbl

\end{document}